\documentclass[12pt]{article}
\usepackage[left=2.50cm, right=2.50cm, top=2.50cm, bottom=2.50cm]{geometry}
\usepackage[utf8]{inputenc}
\usepackage{cancel}
\usepackage[english]{babel}
\usepackage{physics}
\usepackage{hyperref}
\usepackage{amsthm}
\usepackage{mathtools}
\usepackage[normalem]{ulem}
\usepackage{graphicx}
\usepackage{changepage}
\usepackage{amssymb}
\usepackage{verbatim}
\usepackage{empheq}
\usepackage{xcolor}
\usepackage{tikz}
\usepackage{multirow}
\usepackage{amsmath}
\usepackage{centernot}
\usepackage[hang,flushmargin]{footmisc} 
\usetikzlibrary{tikzmark}
\numberwithin{equation}{section}
\hypersetup{hidelinks}

\newlength{\bibitemsep}
\newlength{\bibparskip}
\let\oldthebibliography\thebibliography
\renewcommand\thebibliography[1]{%
  \oldthebibliography{#1}%
  \setlength{\parskip}{\bibitemsep}%
  \setlength{\itemsep}{\bibparskip}%
}
\begin{document}

\par
\bigskip
\Large
\noindent
{\bf 
Notes from the bulk: metric dependence of the edge states of  Chern-Simons theory\\

\par
\rm
\normalsize

\hrule

\vspace{1cm}

\large
\noindent
{\bf Erica Bertolini$^{1,2,a}$}, 
{\bf Giulio Gambuti$^{3,b}$},
{\bf Nicola Maggiore$^{1,2,c}$}\\

\par
\small
\noindent$^1$ Dipartimento di Fisica, Universit\`a di Genova, Italy.
\smallskip

\noindent$^2$ Istituto Nazionale di Fisica Nucleare - Sezione di Genova, Italy.

\smallskip

\noindent$^3$ Rudolf Peierls Centre for Theoretical Physics, Oxford University, U.K.

\smallskip


\vspace{1cm}

\noindent
{\tt Abstract~:}

The abelian Chern-Simons theory is considered on a cylindrical spacetime $\mathbb{R} \times D$, in a not necessarily flat Lorentzian background. As in the flat bulk case with planar boundary, we find that also on the radial boundary of a curved background  a Ka\c{c}-Moody algebra exists, with the same central charge as in the flat case, which henceforth depends neither on the bulk metric nor on the geometry of the boundary. The holographically induced theory on the 2D boundary is topologically protected, in the sense that it describes a Luttinger liquid, no matter which the bulk metric is. The main result of this paper is that 
a remnant of the 3D bulk theory resides in the chiral velocity of the edge modes, which is not a constant like in the flat bulk case, but it is local, depending on the determinant of the induced metric on the boundary. This result may provide a theoretical framework for the recently observed accelerated chiral bosons on the edge of some Hall systems.

\vspace{.5cm}

\vspace{\fill}
\noindent{\tt Keywords:} 
Boundary Quantum Field Theory; Quantum Field Theory On Curved Spacetime; Topological Field Theories; Chern-Simons Theories; Topological States of Matter
\vspace{1cm}

\hrule
\noindent{\tt E-mail:
$^a$erica.bertolini@ge.infn.it,
$^b$giulio.gambuti@physics.ox.ac.uk,\\
$^c$nicola.maggiore@ge.infn.it.
}
\newpage

\section{Introduction}

Topological Field Theories (TFT) represent a paradigmatic example of how  boundaries may be relevant in quantum field theory. In fact TFT are characterized by the defining property of having $global$ observables only, and not $local$. This means that the word ``observable'' in this context is a kind of oxymoron, being 
of geometrical, rather than physical, nature. Examples
are the number of handles or genus of the manifold on which the TFT are built, or the classifications of all possible knots compatible with that particular manifold. These results are quite important, and motivated the great interest in the community of theoretical physicists which arose on TFT after a couple of seminal papers at the end of the eighties of the past century \cite{Witten:1988ze,Witten:1988hf}, where problems typical of mathematics and mathematical physics were for the first time successfully faced by quantum field theory methods. On the other hand, the fact that TFT have vanishing energy-momentum tensor, hence vanishing energy density and, above all, vanishing Hamiltonian, justified as well the colder attitude of the larger part of less formal physicists toward TFT.  The introduction of boundaries in TFT drastically changed this scenario. One of the first, and still more important, examples has been the classification of all known rational Conformal Field Theories (CFTs), obtained by the introduction of a boundary in 3D\footnote{By 3D we mean 2 (space) + 1 (time) dimensions.} topological Chern-Simons (CS) theory \cite{Moore:1989yh}. In that case, the boundary is a circle on a flat bulk 3D manifold. Later, the introduction of a boundary in TFT gave a field theoretical predictive description of the edge modes of the Fractional Quantum Hall Effect (FQHE)\cite{Wen:1992vi,Stone:1990iw,Frohlich:1994vq,Bieri:2010za,Froehlich:2018oce} and of another kind of topological state of matter, the Topological Insulators, both in 3D and 4D \cite{Hasan:2010xy,Qi:2011zya,Hasan:2010hm,moorenature}, 
which are fermionic, despite the fact that the bulk Degrees Of Freedom (DOF) are bosonic \cite{Aratyn:1984jz, Amoretti:2013xya}. In particular, the CS coupling constant has been found to be tightly related to the filling factor of FQHE \cite{Zhang:1992eu} 
and to the central charge of the Ka\c{c}-Moody (KM) algebras \cite{Kac:1967jr,Moody:1966gf} formed by conserved currents on the edge of CS theory \cite{Blasi:1990jq,Emery:1991tf,Balachandran:1991dw}. Similarly, another class of TFT, the BF models \cite{Horowitz:1989ng,Karlhede:1989hz,Birmingham:1991ty,Blasi:2005vf,Blasi:2019wpq}, 
which can be defined on any spacetime dimensions, have been found to describe the bulk theory of the Topological Insulators \cite{Cho:2010rk,Balachandran:1992qg}. 
Hence, the introduction of a boundary in TFT is the only way to give local, physical and measurable observables to theories which otherwise show ``only'' formal interest \cite{Chen:2015gma,Geiller:2019bti,Tiwari:2017wqf}, and major results have been achieved in condensed matter theory, mainly. Particularly interesting is the formulation of a quantum field theory with boundary developed by Symanzik in \cite{Symanzik:1981wd} for the description of the Casimir effect, which is the textbook example of the physical consequences of the presence of a boundary. It is instructive and interesting to remind how a boundary can be introduced {\it \`a la} Symanzik in a field theory: the presence of a boundary manifests itself as a ``separation'' condition on the propagators of the theory, which vanish if computed on points lying on opposite parts of the boundary \cite{Amoretti:2014iza,Amoretti:2013nv}. In Symanzik's case, the bulk theory was not topological, and consequences of the presence of the boundary have been studied also in non-TFT \cite{Maggiore:2018bxr,Maggiore:2019wie,Bertolini:2020hgr,Amoretti:2017xto,Amoretti:2014kba}. One of these, which motivated this work, is the dependence of the boundary physics on the bulk manifold.  This paper is particularly concerned with the consequences of the results to condensed matter theory, where, usually, all measurable quantities, like for instance filling factors and velocities of the edge modes, have been obtained for $flat$ bulk manifolds, with planar, or radial (in the case of CS on a disk \cite{Dunne:1998qy} for instance), boundary. Experiments are pushing theoretical investigations, since recently (see Section 4 for more details)  $accelerated$ chiral bosons have been observed on the edge of some particular Hall systems, which cannot be explained by the usual flat Chern-Simons with boundary paradigm. Similar experimental evidence is currently sought in the other relevant topological states of matter, $i.e.$ the topological insulators. Hopefully, the procedure described in our paper might  be applied to that case as well, using as a bulk theory BF instead of Chern-Simons.
In all the cases we mentioned above, the lower dimensional edge dynamics depends on the bulk only through its parameters, and the details of the bulk manifold are somehow hidden by the particular flat choice. It appears interesting and reasonable to ask if, how and where the details of the metric reveal themselves in the edge observables, to see which are the quantities depending on the bulk metric and, maybe even more interestingly, which are the physical quantities really universal, or topologically protected. The paper is organized as follows. In Section \ref{sec2} the CS theory on a cylindrical spacetime $\mathbb{R} \times D$, in a not necessarily flat Lorentzian background is introduced, and the Equations Of Motion (EOMs), together with the most general Boundary Conditions (BC) are derived. We find the Ward identity, broken by the boundary, and we study the existence of a KM algebra, with constant and positive central charge. In Section \ref{sec3} we derive the 2D theory, holographically induced on the boundary of the 3D CS theory. The holographic contact is imposed, which relates the parameters of the edge theory to the bulk ones. The dependence on the bulk through the induced metric on the boundary is discussed. Our concluding remarks are summarized in Section \ref{sec4}.

\subsection{Notations and conventions}

In this paper we will make use of the following notations, concerning indices and coordinates
\begin{equation}
	\begin{split}
	\mu,\nu,\rho,...=& \{0,1,2\}=\{t,r,\theta\}\\
	i,j,k,...=& \{0,2\}=\{t,\theta\}\ ,
	\end{split}\label{1.1}
\end{equation}
\begin{empheq}{align}
x=&(x_0,x_1,x_2)=(t,r,\theta)\quad\mbox{on the 3D bulk}\label{1.2}\\
X=&(x_0,x_2)=(t,\theta)\quad\mbox{on the 2D boundary }r=R\ .\label{1.3}
\end{empheq}
The Levi-Civita tensor $\epsilon^{\mu\nu\rho}(x)$ is written in terms of the Levi-Civita symbol $\tilde\epsilon^{\mu\nu\rho}$ (taken with the usual normalization $\tilde\epsilon^{012} = 1$) as 
\begin{equation}
\epsilon^{\mu\nu\rho}(x)=\frac{\tilde\epsilon^{\mu\nu\rho}}{\sqrt{-g(x)}}\ ,
\label{1.4}\end{equation}
where $g(x)$ is the determinant of the metric tensor $g_{\mu\nu}(x)$ with Lorentzian signature.
The scalar Dirac delta function $\delta^{(n)}(x-x')$ is written in terms of the flat Dirac delta $\tilde\delta^{(n)}(x-x')$, which is a scalar density, in the following way \cite{Poisson:2011nh}
\begin{equation}
\delta^{(n)}(x-x') = \frac{\tilde\delta^{(n)}(x-x')}{\sqrt{-g}}\ ,
\label{1.5}\end{equation}
with $g\equiv g(x)$, acting on a test function $f(x)$ as follows
\begin{equation}
\int d^nx\,\sqrt{-g}\,\delta^{(n)}(x-x')f(x) = 
\int d^nx\,\tilde\delta^{(n)}(x-x')f(x)
=f(x')\ .
\label{}\end{equation}
Hence, the functional derivative of a generic dual vector field $V_\mu(x)$ is the (1,1) tensor
\begin{equation}
\frac{\delta V_\mu(x)}{\delta V_\nu(x')}=\delta^\nu_\mu\,\delta^{(n)}(x-x')\ .
\label{1.7}
\end{equation}
It is convenient to work in Gaussian Normal Coordinates (GNC), where the line element takes the form
\begin{equation}
ds^2=g_{\mu\nu}(x)dx^\mu dx^\nu =\  dr^2+\gamma_{ij}(x)dX^idX^j\ .
\label{1.9}
\end{equation}
The adoption of the GNC is particularly useful for calculations in situations where one is given a hypersurface $S$, $i.e.$ an $(n-1)$-dimensional embedded submanifold of the n-dimensional manifold $M$ \cite{wald}. In the case treated in this paper, being $M$ diffeomorphic to a cylinder $M \simeq \mathbb{R} \times D$, this choice is the most natural one, since it immediately provides the induced metric on the boundary. The GNC are characterized by the presence of one coordinate (the one normal to the hypersurface, in our case $r$) such that $g_{rr} =1$ and the off-diagonal terms vanish: $g_{rt}=g_{r\theta}=0$ \cite{weinberg}. These are three extra conditions which might be seen (as in \cite{Carroll:2004st} p. 284, and \cite{d'inverno} Eq. (13.4)) as  ``gauge conditions''  on the metric, corresponding to the threefold coordinate freedom $x^\mu\rightarrow x^{\prime\;\mu}= x^{\prime\;\mu}(x)$. Simple examples of GNC are the Cartesian coordinates on Minkowski space, or polar coordinates in Euclidean 2 and 3-space, or the Robertson-Walker coordinates used in cosmology \cite{Carroll:2004st}.  
Using these coordinates, the determinant of the induced metric on the boundary
\begin{equation}
\gamma_{ij}(X)=\frac{\partial x^\mu}{\partial X^i}\frac{\partial x^\nu}{\partial X^j}g_{\mu\nu}(x)\ ,
\label{1.10}
\end{equation}
is (see (D.30) of \cite{Carroll:2004st})
\begin{equation}
\sqrt{-\gamma}=\sqrt{-g}\ ,
\label{1.11}
\end{equation}
which, in particular, holds on the boundary $r=R$. 
Lastly,  in curved spacetime the Heaviside step distribution is a scalar, and its derivative, as in flat spacetime, is (see Appendix \ref{appA})
\begin{equation}
\nabla_\mu\theta(R-r)=-n_\mu
\frac{\sqrt{-\gamma}}{\sqrt{-g}}\delta(R-r)=-\delta_\mu^r\delta(R-r)\ ,
\label{1.12}
\end{equation}
where in the last equality we used the property \eqref{1.11} of GNC, for which the fractional prefactor is equal to one and the unit normal vector is $n_\mu=\delta^r_\mu$.
\section{CS with boundary at $r=R$}\label{sec2}

\subsection{The action}\label{sec2.1}
We consider the abelian 3D CS theory on a cylindrical spacetime $\mathbb{R} \times D$, where the model is confined to the closed subspace $0\leq r\leq R$.  This is achieved by introducing a Heaviside step distribution $\theta(R-r)$ in the action. The CS bulk term is then
\begin{equation}
S_{bulk}
=
\frac{\kappa}{2}\int d^3x\;  \theta(R-r)\; \tilde\epsilon^{\mu\nu\rho}\;A_\mu\partial_\nu A_\rho
\label{2.1}\end{equation}
We remark that although the coupling constant $\kappa$ could be reabsorbed by a redefinition of the fields, it is useful to leave it explicit, in order to keep track of the contributions of the bulk theory to the physics of the boundary. Moreover, since the CS theory, being topological,  has vanishing energy momentum tensor, there is no constraint on the sign of $\kappa$ from requiring a positive energy density. Still, $\kappa$ should be a positive constant, as we shall see later. 
The gauge fixing term is chosen to be
\begin{equation}
S_{gf }
=\int d^3x\; \sqrt{-g}\;  \theta(R-r)\; b\;n^\mu A_\mu\ ,
\label{2.2}\end{equation}
where $n^\mu=(0,1,0)$ is a unit vector. The field $b(x)$ is the Nakanishi-Lautrup  Lagrange multiplier \cite{Nakanishi:1966zz,Lautrup:1967zz} which implements
 the radial gauge choice
\begin{equation}
\frac{\delta S}{\delta b}
=n^\mu A_\mu
=A_r
=0\ .
\label{x2.3}\end{equation}
The choice of the radial gauge-fixing $A_r=0$ is analogous to the choice of the GNC \eqref{1.9}. Both are the most convenient choices in presence of the boundary $r=R$: the latter on the metric $g_{\mu\nu}(x)$ and corresponds to the reparametrization invariance and the former on the gauge field $A_\mu(x)$ and comes from gauge invariance. Both transformations (ordinary gauge symmetry and diffeomorphism invariance) are broken by the boundary. And, in both cases, physical results should not depend on them. For instance, the KM algebra on the boundary of CS theory has been shown to exist, for flat spacetime, both in covariant \cite{Blasi:1990jq} and axial \cite{Emery:1991tf} gauge. We expect that the results contained in this paper can be obtained also without using GNC. We do not know of an analogous calculation done elsewhere, and certainly it would be interesting to check this guess.
External sources are coupled  
to the $A_\mu(x)$ field, through the term 
\begin{equation}
S_{ext}=\int d^3x\; \sqrt{-g}\;   \theta(R-r)\; J^\mu A_\mu\ .
\label{2.3}\end{equation}
In addition, the presence of a boundary induces an extra term in the action, as the most general boundary term compatible with power counting:
\begin{equation}
S_{bd}=\int d^3x\, \sqrt{-g}\,  \delta(r-R)\,\frac{1}{2}\; T^{ij}A_iA_j\ ,
\label{2.4}
\end{equation}
where $T^{ij}=T^{ji}$ is a symmetric matrix. 
Notice that in a curved spacetime all the coefficients appearing in the action may depend on the metric, which is dimensionless, and hence on the coordinates, but only through the metric. In particular, $T^{ij}$ in $S_{bd}$ might depend implicitly on the coordinates $T^{ij}=T^{ij}(\gamma(X))$, where $\gamma(X)$ is the induced metric. An explicit dependence of $T^{ij}$ on the coordinates ``outside'' the induced metric $(T^{ij}=T^{ij}(\gamma(X);X))$ is forbidden, since in the flat limit $T^{ij}$ should reduce to a constant symmetric matrix: $\left.T^{ij}(\gamma(X))\right|_{\gamma^{ij}=\eta^{ij}}=T^{ij}$.
Concerning the lower-dimensional term $S_{bd}$ \eqref{2.4}, we follow \cite{Symanzik:1981wd}, where Symanzik added an analogous term to the bulk action in order to implement the boundary conditions by means of the action principle. While Symanzik imposes boundary conditions and then sees which is the boundary term in the action which implements them, we generalize that approach not imposing any boundary condition {\it a priori}, but 
writing the most general boundary term, compatible with locality and power counting, to find out which are the most general boundary conditions compatible with it.
A similar boundary term appears also in \cite{Elitzur:1989nr} (Eq. (3.2)) for CS theory. In both cases, the main reason is to provide, by means of the modified equations of motion, the most general boundary conditions which need to be fixed. The introduction of such a local functional allows to introduce the boundary conditions in a systematic way by means of a variational principle, and not by hand. Another way to see this, is to think about what happens in General Relativity, where the Einstein-Hilbert action is not invariant under general coordinate transformations, and the Gibbons-Hawking term needs to be added. In a similar way, here the presence of the boundary spoils the gauge invariance of the CS action. The boundary term must only satisfy the general requirements of power counting and locality. Gauge invariance and/or residual 2D covariance must not be required on it, otherwise, one would not recover the boundary dynamics which characterizes TFT (see for instance \cite{Elitzur:1989nr}, where the prescription of introducing a non covariant boundary term $\propto \int_{\partial Y} Tr A_1A_2$ is adopted in order to get consistent boundary conditions). Under this respect, an analogy can be done with massive gravity, where non-Lorentz invariant terms are proposed in order to generalize the standard Fierz-Pauli theory \cite{Rubakov:2008nh, Blasi:2017pkk, Blasi:2015lrg}.
Finally, the total action of the theory, considering the bulk, gauge-fixing, external source and boundary terms, is
\begin{equation}
S=S_{bulk} +S_{gf} + S_{ext} + S_{bd}\ .
\label{2.6}\end{equation}

\subsection{Equations of motion and boundary conditions}\label{sec2.2}

From the action $S$ \eqref{2.6} we get the EOM
\begin{equation}\label{2.7}
\frac{\delta S}{\delta A_\lambda}=\theta(R-r)\left(\kappa\epsilon^{\lambda\mu\nu}\partial_\mu A_\nu+b\;n^\lambda+J^\lambda\right)+\delta(R-r)\delta^\lambda_j\left(\frac{\kappa}{2}\epsilon^{i1j}+T^{ij}\right)A_i=0\ .
\end{equation}
Applying the operator 
$\lim_{\epsilon\to R}\int_{\epsilon}^{R}dr$ to the EOM \eqref{2.7}, the following  BC can be derived
\begin{equation}
\left. \left(\frac{\kappa}{2}\frac{\tilde\epsilon^{1ij}}{\sqrt{-g}}+T^{ij}\right)A_j \right|_{r=R}= 0\ .
\label{2.9}\end{equation}
In a more compact way, the BC can be written in matricial form
\begin{equation}
M^{ij}A_j=0\ ,
\label{2.12}\end{equation}
where
\begin{equation}
M=\left(
\begin{array}{cc}
c_1 & c_2-\tilde\kappa \\
c_2+\tilde\kappa & c_3
\end{array}
\right)\ ,
\label{2.13}\end{equation}
having defined
\begin{equation}
c_1\equiv T^{00}\ ;\ c_2\equiv T^{02}=T^{20} \ ; \ c_3\equiv T^{22}
\label{2.14}
\end{equation}
and 
\begin{equation}
\tilde\kappa\equiv\frac{\kappa}{2}\frac{\tilde\epsilon^{012}}{\sqrt{-g}}\ .
\label{2.15}
\end{equation}
Even though in \eqref{2.15} the value of the 012-component of the Levi-Civita symbol is $\tilde\epsilon^{012}=1$, we choose to keep it explicit, to enhance the 
fact that $\tilde\kappa(X)$ is a scalar, and not a scalar density,  as it would appear by hiding $\tilde\epsilon^{012}$.
The most general solution of the BC \eqref{2.12} which does not involve vanishing components of the gauge field is  
\begin{equation}
A_t + v A_\theta=0\ ,
\label{2.16}\end{equation}
where the coefficient $v(X)$  in \eqref{2.16} must be different from zero and is not constant, since it depends on the parameters appearing in the action $S$ \eqref{2.6} according to the following relations
\begin{equation}
c_1= 0 \ ;\ c_2 = \tilde\kappa\ ;\  c_3\neq 0\ ;\  2\tilde\kappa A_t+c_3A_\theta=0\label{2.17}\ \Rightarrow\ v=\frac{c_3}{2\tilde\kappa}\ ;
\end{equation}
\begin{equation}
c_1\neq 0\ ;\ c_2=-\tilde\kappa\ ;\ c_3=0\ ;\ c_1A_t-2\tilde\kappa A_\theta=0\label{2.18}\ \Rightarrow\ v=-\frac{2\tilde\kappa}{c_1}\ ;
\end{equation}
\begin{equation}\label{2.19}
c_1\neq 0\ ;\ 
c_2\neq\pm\tilde\kappa\ ;\
c_3\neq0\ ;\
c_1c_3-c_2^2+\tilde\kappa^2=0\ ;\
\left\{
\begin{array}{ll}
c_1A_t+(c_2-\tilde\kappa)A_\theta=0\ &\Rightarrow\ v=\frac{c_2-\tilde\kappa}{c_1}\\
\mbox{or}&\\
(c_2+\tilde\kappa)A_t+c_3A_\theta=0\ &\Rightarrow\ v=\frac{c_3}{c_2+\tilde\kappa}\ .
\end{array}
\right. 
\end{equation}
The parameter $v$ appearing in the boundary condition \eqref{2.16} depends on $c_i$ \eqref{2.14}, which are the components of the boundary parameter $T^{ij}$ appearing in $S_{bd}$ \eqref{2.4}, and on $\tilde\kappa$ \eqref{2.15}. Hence, $v$ may depend on the coordinates through the induced metric $\gamma^{ij}$  and on its determinant $\gamma$ (which in GNC is equal to $g$). No explicit dependence on the coordinates $X=\{t,\theta\}$ is possible because, as we said, in the flat limit $T^{ij}$ should be constant. The boundary condition \eqref{2.16} is of the same type of that derived in \cite{Geiller:2017xad} by an action principle, in analogy to our approach. The difference is that in \cite{Geiller:2017xad}
the analogue of the matrix $T^{ij}$ is constant, which can be obtained after a rescaling by $\sqrt{-g}$. As a consequence, the parameter $v$ appearing in the boundary condition in \cite{Geiller:2017xad}  is constant. In other words, it is always possible to rescale $v$ to a constant by means of a coordinate choice. The main goal of this paper, as we shall see, is that we will be able to relate the parameter $v$ to a measurable quantity, which is the velocity of the chiral edge modes. For this reason we do not rescale $v$ to an arbitrary constant value, but we let $v$ to be determined by a phenomenological input. It is important not to rescale $v$ to a constant value in order to be able to account for $accelerated$ chiral bosons on the edge of certain Hall systems, as we shall see, which are not explained by a constant $v$. 
This is an important point, which concerns two different perspectives. Our approach is the same to the abelian CS description of the Hall systems. From a pure field theoretical point of view, abelian CS model is a free theory, with no coupling constant, for the simple reason that it can be reabsorbed by a rescaling of the gauge field. It is only the nonabelian extension of CS theory which displays a true coupling constant. In condensed matter theory the perspective is different, somehow opposite. The coupling constants are not rescaled by field redefinitions, but are fixed by external phenomenological inputs. This reflects even in a different terminology. For instance, the abelian CS ``coupling constant'', which is oxymoric in field theory, is often called ``CS level'' by a condensed matter - oriented reader, because it is fixed by its relation to the filling factor of Landau levels: $\kappa=\frac{1}{2\pi\nu}$. In curved spacetime the metric plays the role of a dynamical field, which we treat exactly in the same way: we do not rescale the metric by choosing a particular coordinates set, but we let the parameter $v$ to be constrained by the observed chiral velocities of the edge modes. We have seen that $v$ depends on the induced metric and its determinant, and hence we relate the induced metric (which depends on the bulk metric of CS theory) to an observable quantity. And this is the only way, so far, to take into account the observed accelerated chiral velocities, which are not explained by a flat background of CS theory alone. If we were not interested in a phenomenological interpretation of our model, we would rescale $\kappa$ to one and $v$ to whatever value, including zero, which would correspond to a Dirichlet condition on one component of the gauge field. 

\subsection{Ward identity}\label{sec2.3}

The covariant derivative of the EOM \eqref{2.7} is
\begin{equation}\label{2.20}
\nabla_\lambda\frac{\delta S}{\delta A_\lambda}=\theta(R-r)\left[\kappa\epsilon^{\lambda\mu\nu}\nabla_\lambda\nabla_\mu A_\nu+\nabla_\lambda\left(b\;n^\lambda\right)+\nabla_\lambda J^\lambda\right]-\delta(R-r)\left(\kappa\epsilon^{1kj}\partial_kA_j+b+J^r\right)=0\ ,
\end{equation}
where the BC \eqref{2.9} have been used to cancel the $\delta(R-r)$ term on the r.h.s. of \eqref{2.7}.
Noting that
\begin{equation}
\epsilon^{\mu\nu\rho}\nabla_\mu\nabla_\nu A_\rho=0\;,
\end{equation}
we find
\begin{equation}\label{2.23}
\theta(R-r)\nabla_\lambda\left( b\;n^\lambda+J^\lambda\right)-\delta(R-r)\left(\kappa\epsilon^{1kj}\partial_kA_j+b+J^r\right)=0\ .
\end{equation}
Keeping in mind that the covariant divergence is
\begin{equation}\label{2.23'}
\nabla_\mu V^\mu=\frac{1}{\sqrt{-g}}\partial_\mu\left(V^\mu\sqrt{-g}\right)\quad;\quad\nabla_k V^k=\frac{1}{\sqrt{-g}}\partial_k\left(V^k\sqrt{-g}\right)\ ,
\end{equation}
where we used \eqref{1.11} in the second identity, multiplying \eqref{2.23} by $\sqrt{-g}$ and integrating
along the coordinate normal to the boundary, we get\footnote{Notice that in general for an integration along $r$ we should use the invariant measure $\sqrt{g_{rr}}\; dr$ in order to preserve the transformation properties under diffeomorphisms,  but $g_{rr} = 1$ in GNG \eqref{1.9}.}
\begin{eqnarray}
0&=&
\int^\infty_0dr\;\theta(R-r)\left[\partial_r (b\sqrt{-g})+\partial_\lambda\left(J^\lambda\sqrt{-g}\right)\right]
-\int_0^\infty dr\;\delta(R-r)\left[\kappa\tilde\epsilon^{1kj}\partial_kA_j+\left(b+J^r\right)\sqrt{-g}\right]\label{2.24}\nonumber\\
&=& \int^\infty_0dr\;\left\{\partial_r \left[\theta(R-r)b\sqrt{-g}\right]+\bcancel{\delta(R-r)b\sqrt{-g}}+\partial_\lambda\left[\theta(R-r)J^\lambda\sqrt{-g}\right]+\cancel{\delta(R-r)J^r\sqrt{-g}}\right\}\nonumber\\
&&\quad\quad-\int_0^\infty dr\;\delta(R-r)\left[\kappa\tilde\epsilon^{1kj}\partial_kA_j+\left(\bcancel b+\cancel{J^r}\right)\sqrt{-g}\right]\\
&=&\int^\infty_0dr\;\left[\theta(R-r)\partial_k\left(J^k\sqrt{-g}\right)-\delta(R-r)\kappa\tilde\epsilon^{1kj}\partial_kA_j\right]-\left[\left(b+J^r\right)\sqrt{-g}\right]_{r=0}\nonumber\\
&=&\int^\infty_0dr\;\sqrt{-g}\left[\theta(R-r)\nabla_kJ^k-\delta(R-r)\kappa\epsilon^{1kj}\partial_kA_j\right]\ ,\nonumber
\end{eqnarray}
where in the second line we integrated by parts and in the third line we used the fact that, evaluating the EOM \eqref{2.7} for $\lambda=r$ and then going at $r=0$,  we have $\left[b(x)+J^r(x)\right]_{r=0}=0$. 
In fact, \eqref{2.7} at $r=0$ ($\lambda=r$) reads
\begin{equation}\label{b=0}
\left[b+J^r - 2\tilde\kappa \,  \Big( \partial_\theta A_t  - \partial_t A_\theta \Big)\right]_{r=0} = 0\ ,
\end{equation}
and we notice that $A_\theta(x)$ necessarily vanishes at $r=0$, together with its time derivatives, while $A_t(x)$, which in principle might not vanish at $r=0$, at the origin must have vanishing angular derivatives, in order to be well defined,  hence the result. 
Therefore,  the contribution involving the Lagrange multiplier and $J^r$ cancel 
 out, leaving only
\begin{equation}
\int_0^Rdr\; \sqrt{-g}\nabla_kJ^k =
\left.\kappa\tilde\epsilon^{1kj}\partial_kA_j
\right|_{r=R}\ .
\label{2.25}
\end{equation}
Eq.\eqref{2.25} is the Ward identity of the theory, broken at its r.h.s. by the presence of the boundary. As we shall see, it will be crucial for the determination of the boundary algebra and of the 2D theory holographically induced on the boundary. Notice that it holds for any bulk metric, and it is simply the curved extension of its flat counterpart \cite{Maggiore:2017vjf}. From \eqref{2.25}, at vanishing external sources $J^k(x)=0$ ($i.e.$ going on-shell), we find
\begin{equation}
\left.\epsilon^{1kj}\partial_kA_j\right|_{r=R;J=0}=0\ ,
\label{2.26}\end{equation}
which describes a conserved current on the closed boundary $r=R$, whose most general solution is \cite{Nash:1983cq,Warner}
\begin{equation}
A_i(X)=\partial_i\Phi (X) + \delta_{i2}\; C\ ,
\label{2.27}\end{equation}
where $C$ is a constant (which we will show shortly being equal to zero), and $\Phi(X)$ is a scalar field that will play the role of DOF of the 2D boundary theory. The components of the gauge field on the boundary are then
\begin{equation}
A_t (t,\theta)=\partial_t\Phi(t,\theta)\ \ ;\ \
A_\theta(t,\theta)=\partial_\theta\Phi(t,\theta)+C\ .
\label{2.28}\end{equation}
Since we are considering a closed boundary, we also have to impose periodicity conditions on the fields~:
\begin{equation}
A_i(t,\theta)=A_i(t,\theta+2\pi ) \Rightarrow \Phi(t,\theta)=\Phi(t,\theta+2\pi )\ .
\label{2.29}\end{equation}
The value of the constant $C$ in \eqref{2.27} is found by applying the mean value theorem for holomorphic functions, which states that if $f$ is analytic in a region $D$, and $a\in D$, then $f(a)=\frac{1}{2\pi}\oint_{{\cal C}(a)} f$, where ${\cal C}(a)$ is a circle centered in $a$. 
In our case (3D) taking  for $\cal C$ the circular boundary $r=R$ centered at $r=0$ allows us to write
\begin{equation}
\oint_{ring\; R} A_\theta(x) 
=\cancel{\left.\Phi(t,\theta)\right|_{\theta=0}^{\theta=2\pi }} +
 2\pi \; C\ ,
\label{2.30}\end{equation}
where we used the periodicity condition of the boundary field $\Phi(X)$ \eqref{2.29}. This is the value of the bulk field at the center of the ring, and using the fact that  $A_{\theta}(t,r=0,\theta)=0$, we finally get
\begin{equation}
C=0\ .
\label{2.31}\end{equation}
\subsection{Algebra }\label{sec2.4}

The generating functional of the connected Green functions $Z_c[J]$ is defined, as usual, in the following way
\begin{equation}
e^{iZ_c[J]}=\int DA\;Db\;e^{iS[A,b;J]}\  ,
\label{2.32}
\end{equation}
where $S$ is the total action \eqref{2.6}. From \eqref{2.32} we get the 1- and 2-points Green functions
\begin{empheq}{align}
\left.\frac{\delta Z_c[J]}{\delta J^i(x)}\right|_{J=0}=&\;\langle A_i(x)\rangle\label{2.33}\\
\left.\frac{\delta^{(2)} Z_c[J]}{\delta J^i(x)\delta J^j(x')}\right|_{J=0}\equiv&\Delta_{ij}(x,x')=i\langle T(A_i(x)A_j(x'))\rangle\ ,\label{2.34}
\end{empheq}
where the time-ordered product is defined as
\begin{equation}
\langle T(A_j(x)A_l(x'))\rangle\equiv\theta(t-t')\langle A_j(x)A_l(x') \rangle +
\theta(t'-t)\langle A_l(x') A_j(x) \rangle\ .
\label{2.37}
\end{equation}
The algebra is obtained by making the functional derivative with respect to $J^l(x')$ of the Ward identity \eqref{2.25}, and then going on-shell, $i.e$ putting $J=0$~:
\begin{equation}
\frac{ \delta}{\delta J^l(x')} \int_0^R dr\; \sqrt{-g}\;\nabla_k^x\,J^k(x)=
\kappa\; \tilde\epsilon^{1kj}\; \partial^x_k 
\left. 
\frac{\delta^{(2)} Z_c}{\delta J^l(x') \delta J^j(x) }
\right |_{r=R;J=0}\ .
\label{2.35}\end{equation}
Therefore, the first step is to compute
	\begin{equation}\label{2.35'}
		\frac{\delta}{\delta J^l(x')}\nabla^x_k J^k(x)=\frac{\delta}{\delta J^l(x')}\left[\frac{1}{\sqrt{-g}}\partial^x_k\left(J^k\sqrt{-g}\right)\right]
		=\frac{1}{\sqrt{-g}}\partial^x_l\tilde\delta^{(3)}(x-x')\ ,
	\end{equation}
where we used the relation \eqref{1.5} betweeen the scalar and density delta function and the fact of working in GNC \eqref{1.11}.
Using \eqref{2.35'} on the l.h.s.  of \eqref{2.35}, we obtain 
\begin{equation}
\partial_l\tilde\delta^{(2)}(X-X') = 
i\kappa \left.\tilde\epsilon^{1kj}\partial^x_k\langle T(A_j(x)A_l(x'))\rangle\right |_{r=R}\ .
\label{2.36}\end{equation}
Choosing $l=\theta$ in \eqref{2.36} we have~:
\begin{eqnarray}
\partial_\theta\tilde\delta^{(2)}(X-X')
&=&
i \kappa \tilde\epsilon^{1kj} \partial_k^X 
 \left[
 \theta(t-t') \langle A_j(X)A_\theta(X')\rangle +
 \theta(t'-t) \langle A_\theta(X')A_j(X)\rangle
 \right]\nonumber\\
 &=&
 i\kappa \tilde\epsilon^{102} \left(\partial_t\theta(t-t')\right)
 \left(
 \langle A_{\theta}(X)A_\theta(X')\rangle -
 \langle A_\theta(X')A_{\theta}(X)\rangle
 \right) \label{2.38}
 \\
 && +\kappa \tilde\epsilon^{1kj} \left[
i \theta(t-t') \langle \bcancel{\partial_k^XA_j(X)}A_\theta(X')\rangle +
 i\theta(t'-t) \langle A_\theta(X')\bcancel{\partial_k^XA_j(X)}\rangle
 \right] \nonumber\ ,
\end{eqnarray}
where we used the on-shell condition \eqref{2.26}. By defining
\begin{equation}
\left[A_j(X),A_\theta(X')\right]
\equiv
\langle A_j(X)A_\theta(X')\rangle -
 \langle A_\theta(X')A_j(X)\rangle\ ,
\label{2.39}\end{equation}
we get
\begin{equation}
\partial_\theta\tilde\delta^{(2)}(X-X')
=
-i\kappa\tilde\epsilon^{012} \left(\partial_t\theta(t-t')\right) \left[A_\theta(X),A_\theta(X')\right] \ .
\label{2.40}\end{equation}

Finally, integrating over time we find the equal time commutator
\begin{equation}
\tilde\epsilon^{012}\left.\left[A_\theta(X),A_\theta(X')\right]\right|_{t=t'} =
\frac{i}{\kappa}\,
\partial_\theta\tilde\delta(\theta-\theta')\ ,
\label{2.43}\end{equation}
By applying the same procedure to the case $l=t$ of \eqref{2.36}, we find the equal time commutator
\begin{equation}
\left.\left[A_\theta(X),A_t(X')\right]\right|_{t=t'}=0\ .
\label{2.45}\end{equation}
Eq.\eqref{2.43} represents an abelian KM algebra
identical to the one found in the case of CS theory with planar boundary in flat space. We see that the existence of a KM algebra and its central charge depends neither on the bulk metric nor on the details of the boundary. The central charge 
\begin{equation}
c=\frac{1}{\kappa}
\label{2.49}\end{equation}
must be positive (for the unitarity of the associated CFT {\cite{mack,Becchi:1988nh}), and for this reason the coupling constant of the CS theory has to be positive as well
\begin{equation}
\kappa>0\ .
\label{2.50}\end{equation}
Notice that this algebraic method is the only way to determine the sign of the CS coupling constant, since this theory, being topological, has vanishing stress-energy tensor, and hence the usual argument based on the positivity of the energy density cannot be applied here.

\section{2D boundary theory}\label{sec3}

\subsection{The 2D action}

In Section \ref{sec2.3}, we identified the solution of the conserved current equation \eqref{2.26} as the boundary DOF. This will allow us to find the 2D dynamics on the boundary, in fact we can express the equal time commutation relation \eqref{2.43} in terms of the boundary field $\Phi(X)$, by using \eqref{2.28}
\begin{equation}\label{3.1}
[\partial_\theta\Phi(X),\tilde\epsilon^{012}\partial_{\theta'}\Phi(X')]=i\frac{1}{\kappa}\;\partial_\theta\tilde\delta(\theta-\theta')\ ,
\end{equation}
which, rewritten as
\begin{equation}\label{3.2}
[\Phi(X), \kappa\tilde\epsilon^{012}\partial_{\theta'}\Phi(X')]=i\;\tilde\delta(\theta-\theta')\ ,
\end{equation}
can be interpreted as a canonical commutation relation in curved spacetime
\begin{equation}\label{3.3}
[q(t,x),p(t,x')]=i\;\tilde\delta(x-x')\ ,
\end{equation}
with $p(t,x)$ a density (see (9.87-90-91) of \cite{Carroll:2004st}), provided that we match the canonical variables of the 2D theory as follows
\begin{equation}\label{3.4}
q(X)\equiv\Phi(X)\quad;\quad p(X)\equiv \tilde\epsilon^{012}\kappa\partial_\theta\Phi(X)\ .
\end{equation}
Here again, the presence of the $\tilde\epsilon^{012}$ factor in \eqref{3.4} is of great importance, since it makes $p(X)$ a scalar density, according to the standard definition {\cite{Basler:1991st}. Once identified the canonical variables, we can begin the search for the induced boundary theory, by  writing the most general  Lagrangian compatible with power counting. Noting that the scalar field has vanishing mass dimension, we find
\begin{equation}
\mathcal{L}_{2D}=\sqrt{-\gamma}\;
(a^{ij}\partial_i\Phi\partial_j\Phi+b^i\partial_i\Phi+c)\ ,
\label{3.5}
\end{equation}
where the coefficients $a^{ij},\ b^i,\ c$
\begin{enumerate}
\item must be tensor quantities, $i.e.$ symmetric tensor of rank (2,0), contravariant vector and scalar respectively, in order that the Lagrangian \eqref{3.5} is a scalar density\; ;
\item must have mass dimension 0, 1 and 2, respectively\; ;
\item may depend on the scalar field $\Phi(X)$, the metric $\gamma_{ij}(X)$ and/or its determinant $\gamma(X)$ (but not on their derivatives), since in 2D the scalar field $\Phi(X)$, like the metric, is dimensionless, and thus the power counting is preserved. In other terms, the 2D action \eqref{3.5} is written as a derivative expansion.
\end{enumerate}
We also notice that the scalar field $\Phi(X)$ is defined by means of \eqref{2.28} up to a shift transformation
\begin{equation}
\delta_{shift}\Phi=\alpha\ ,
\label{3.6}
\end{equation}
with $\alpha$ constant, which implies that the action 
\begin{equation}
S_{2D}=\int d^2X\; {\cal L}_{2D}
\label{3.7}\end{equation}
describing the boundary theory should possess the same symmetry as well, $i.e.$
\begin{equation}
 \delta_{shift}S_{2D}=0\;.
 \label{3.8}\end{equation}
An immediate consequence of the shift symmetry \eqref{3.8} is that $a^{ij}$ and $c$ must be constant with respect to $\Phi$. For what concerns $b^i$, it is easily seen that it may admit a linear dependence on the scalar field. In fact, if $b^i=b_1^i+b_2^i\Phi$, where $b_1^i$ and $b_2^i$ do not depend on $\Phi$, but may depend on the induced metric $\gamma_{ij}$ and on its determinant $\gamma$, we have that 
\begin{equation}
\delta_{shift}\int d^2X\; \sqrt{-\gamma} \;
b^i\partial_i\Phi=
\int d^2X\; \sqrt{-\gamma} \;
\frac{\partial b^i}{\partial\Phi}\alpha\partial_i\Phi=
\alpha\int d^2X\; \sqrt{-\gamma} \;\nabla_i\left(\frac{\partial b^i}{\partial\Phi}\Phi\right)\ ,
\label{linearb}\end{equation}
which is a vanishing boundary term. 
Therefore, the corresponding term in the 2D Lagrangian \eqref{3.5} does not contribute, being a boundary term as well.
\begin{equation}
\int d^2X\; \sqrt{-\gamma} \;
b^i\partial_i\Phi=
\int d^2X\; \sqrt{-\gamma} \;
(b_1^i\partial_i\Phi+b_2^i\Phi\partial_i\Phi)=
\int d^2X\; \sqrt{-\gamma} \;\nabla_i
(b_1^i\Phi+\frac{1}{2}b_2^i\Phi^2)\ .
\end{equation}
The most general 2D action satisfying the shift symmetry  is 
 \begin{equation}
S_{2D} = \int d^2X\; \sqrt{-\gamma} \;
	(a^{ij}\partial_i\Phi\partial_j\Phi+c) =
\int d^2X\; \sqrt{-\gamma}\;\left(a_0\partial_t\Phi\partial_t\Phi+2a_1\partial_t\Phi\partial_\theta\Phi+a_2\partial_\theta\Phi\partial_\theta\Phi +c \right),
\label{3.10}
\end{equation}
where $a_0\equiv a^{00}$, $a_1\equiv a^{02}$ and $a_2\equiv a^{22}$. The last term does not contribute to the equations of motion of the scalar field $\Phi$ and will be omitted. The most general dependence of the dimensionless $a^{ij}$ on the induced metric $\gamma^{ij}$ is
\begin{equation}\label{3.10'}
a^{ij}=\alpha^{ij}(\gamma)+\beta(\gamma)\;\gamma^{ij}\ ,
\end{equation}
where $\alpha^{ij}(\gamma)$ and $\beta(\gamma)$ may depend at most on the metric determinant. What is left now is to verify under which conditions the Lagrangian ${\cal L}_{2D}$ is compatible with the canonical variables identified from the KM algebra through \eqref{3.4}. This is reached by requiring
\begin{equation}\label{3.11}
\frac{\partial\mathcal{L}_{2D}}{\partial\dot q}=p=\tilde\epsilon^{012}\kappa\partial_\theta\Phi\ .
\end{equation}
The l.h.s.  of \eqref{3.11} is
\begin{equation}\label{3.12}
\frac{\partial\mathcal{L}_{2D}}{\partial\dot q}=\frac{\partial\mathcal{L}_{2D}}{\partial(\partial_t\Phi)}=\sqrt{-\gamma}(2a_0\partial_t\Phi+2a_1\partial_\theta\Phi)\ ,
\end{equation}
therefore we must ask
\begin{equation}
a_0=0\quad;\quad a_1=\tilde\kappa 
\quad\quad
\; (a_2\ \mbox{free})\ ,\label{3.13}
\end{equation}
where $\tilde\kappa$ is the scalar function related to the CS coupling constant through \eqref{2.15}. Eq. \eqref{3.13} represents a constraint on the metric dependence of $a^{ij}$ in \eqref{3.10'}, in fact it must be
\begin{empheq}{align}
0&=a_0=a^{00}=\alpha^{00}+\beta\;\gamma^{00}\\
\tilde\kappa&=a_1=a^{02}=\alpha^{02}+\beta\;\gamma^{02}\ .
\end{empheq}
Since we do not want to impose any unnecessary condition on the components of the induced metric $\gamma^{ij}$, that leads to the request
\begin{equation}
\beta=0\ ,
\end{equation}
therefore
\begin{equation}
a_2=a^{22}=\alpha^{22}(\gamma)\ ,
\label{a2}\end{equation}
which means that $a_2$ is a free parameter depending at most on the metric determinant: $a_2=a_2(\gamma)$. Hence, up to terms which do not contribute to the equation of motion of the scalar field $\Phi$, the action $S_{2D}$ \eqref{3.7} is 
\begin{equation}\label{3.14}
S_{2D}=\int d^2X\sqrt{-\gamma}\;\left(
\tilde\kappa\partial_t\Phi
+a_2\partial_\theta\Phi
\right)\partial_\theta\Phi \ .
\end{equation}

\subsection{Holographic contact}

From the action $S_{2D}$ \eqref{3.14} we get the following EOM\\
\begin{equation}\label{3.15}
\partial_\theta\left(
\partial_t\Phi+\frac{a_2}{\tilde\kappa}\;\partial_\theta\Phi\right)=0\;,
\end{equation}
where we divided by the CS coupling constant $\kappa$. Our task is now to find under which conditions the BC \eqref{2.16}, written in terms of the boundary field $\Phi(X)$ by means of \eqref{2.28} (with $C=0$ \eqref{2.31})
\begin{equation}
\partial_t\Phi+v\,\partial_\theta\Phi=0\ ,
\label{3.16}
\end{equation}
can be related to the EOM \eqref{3.15}. 
The BC \eqref{3.16}
is the equation of a chiral boson whose velocity $v$, defined by \eqref{2.16}, may depend on the induced metric $\gamma^{ij}$ through the coefficients $T^{ij}$ appearing in $S_{bd}$ \eqref{2.4}. We recover here the physical interpretation of the parameter $v$ appearing in the boundary conditions \eqref{2.16} as the chiral velocity of the edge modes living on the boundary of CS theory, which are known, in the framework of FQHE, to be measurable quantities. In most Hall systems, the observed chiral velocity is constant, and this is achieved by a CS theory built on a flat 3D spacetime with planar boundary \cite{Maggiore:2017vjf}. The novelty we are finding here, is that $v$ is now a $local$ quantity, in particular depending on time, which is a consequence of considering the CS theory on a curved, instead of flat, spacetime. This corresponds to $accelerated$ edge chiral bosons, which have been indeed recently observed (see Section 4 for more details), and which cannot be explained by putting a boundary on a flat background. The fact that $v$ is a phenomenological parameter requires that it should be determined by experimental inputs, hence a choice of coordinates, which would set $v$ to an arbitrary, possibly constant, or even vanishing, value, should not be done $a\ priori$. We might rather say that the $right$ choice of coordinates is that of the laboratory frame where the chiral accelerated velocity is measured. 
Compatibility between the BC \eqref{3.16} and the EOM \eqref{3.15} is reached if we require that EOM$\equiv\partial_\theta$BC, or introducing, as in \cite{Geiller:2017xad}, the field $n(t,\theta)\equiv\frac{1}{2\pi}\partial_\theta\Phi(t,\theta)$. In any case, the following condition must hold
\begin{equation}\label{3.17}
v=\frac{a_2}{\tilde\kappa}\ .
\end{equation}
Eq.\eqref{3.17} relates the 
parameters of the two theories, thus establishing a (holographic) link between the bulk ($v,\tilde\kappa$) and the induced boundary ($a_2$). 
From \eqref{3.17} we see that, because of \eqref{a2}, the holographic contact makes $v$ depend only on the determinant $\gamma$ of the induced metric $\gamma_{ij}$, and not on its components. This constitutes a restriction on all possible $v$ appearing in the BC \eqref{2.16}, which, instead, might depend on the determinant of the induced metric $and$ on its components as well. This constraint, however, does not spoil the possible local character of $v$.
Eq. \eqref{3.17} is a condition on the parameters which appears in $S_{bd}$ for the holographic bulk-boundary contact to be possible. Therefore the edge velocity depends on the boundary conditions for the Chern-Simons gauge field.
The issue of the determination of the chiral velocities has been discussed in details by Wen (in \cite{Wen:1989mw} and \cite{Wen:1990qp} for instance), who remarked that it cannot be determined from the bulk action (nor be rescaled to an arbitrary value), and that it is thus appropriate to take the velocity $v$ as a phenomenological parameter. This is the phenomenological counterpart of the considerations made in \cite{Frohlich:1990xz}: CS theory provides merely an effective large-distance description which captures the general symmetries of the manybody theory and has, as such, intrinsic limitations. For instance, it should not be expected to capture certain details such as the velocity of the edge chiral bosons. In \cite{Frohlich:1990xz} it is correctly argued that
the  determination of the chiral velocity should come from  gauge-breaking terms present in the full Lagrangian, which, however cannot be identified with the gauge fixing term, since physical results should not depend on the gauge choice, hopefully. In our approach, it is clear that gauge-breaking term present in the full Lagrangian thanks to which the TFT acquires local degrees of freedom is $S_{bd}$ (2.5). The chiral velocity $v$ depends on the boundary condition \eqref{2.16} which, in turn, depends on the parameters appearing in $S_{bd}$. It is the breaking of the gauge symmetry which starts the game: the breaking of the gauge symmetry due to the boundary manifests itself in the breaking of the Ward identity \eqref{2.25}, which yields the irrotationality condition \eqref{2.26}, which identifies the local bosonic boundary degrees of freedom. It is from the broken Ward identity that the algebra \eqref{2.43} is derived, and then all the rest follows, as described in this paper. None of these results depend on the particular gauge choice, as we already remarked.
The 2D action \eqref{3.14} can now be entirely written  in terms of the parameters appearing in the 3D bulk theory \eqref{2.6} as follows
 \begin{equation}
S_{2D}=\frac{\kappa}{2}\int d^2X\; \tilde\epsilon^{012}\;
\left(\partial_t\Phi+v\,\partial_\theta\Phi\right)\partial_\theta\Phi\ ,
\label{3.18}
\end{equation}
where $\kappa$ is the CS coupling constant and $v(t,\theta)$, appearing in the BC \eqref{2.16}, depends on the parameters of the boundary action $S_{bd}$ \eqref{2.4} and, through $\tilde\kappa$, on the determinant of the induced metric $\gamma(t,\theta)$.
We observe that the only dependence of the action $S_{2D}$ on the bulk metric is concealed in $v(t,\theta)$. 
We might say that, in this sense, the metric form of \eqref{3.18} is protected, and it would be interesting to study wether this property holds in higher dimensions. Physically, the 2D action \eqref{3.18} can be immediately identified with the Luttinger theory \cite{Haldane:1981zza}, relevant example of the bosonization phenomenon, for which the density operator $n(t,\theta)$, written in terms of chiral fermions, is
\begin{equation}
n(t,\theta)=:\psi^\dagger(t,\theta)\,\psi(t,\theta):\ .
\label{3.19}
\end{equation}
Bosonization, $i.e.$ fermion/boson correspondence, is achieved through the identification
\begin{equation}
n(t,\theta)=\frac{1}{2\pi}\partial_\theta\Phi(t,\theta)\ ,
\label{3.20}
\end{equation}
where $\Phi(X)$ is just the chiral boson \eqref{3.16} found as the edge state of the CS theory. The density operator \eqref{3.19} satisfies the following commutation relation
\begin{equation}
[n(t,\theta),n(t,\theta')]=i\frac{\nu}{2\pi}\partial_\theta\delta(\theta-\theta')\ ,
\label{3.21}
\end{equation}
where $\nu$ is the filling factor of the FQHE. Eq.\eqref{3.21}, with the bosonisation relation \eqref{3.20},
is exactly the KM algebra we found in \eqref{3.1}, by means of which
we can physically interpret the field $A_\theta(t,\theta)$ as the density operator $n(t,\theta)$ \eqref{3.19}. 
We also notice how the boundary theory makes sense for $any$ dependence of $v$ on $X$. Indeed, from the perspective of \eqref{3.16}, the EOM derived from the action \eqref{3.14} reads
\begin{equation}
\partial_tn+\partial_\theta(vn)=0\ ,
\end{equation}
which represents the continuity equation of a 1D fluid of velocity $v$ and density $n$, as already remarked in the case of flat bulk metric in \cite{Blasi:2008gt}. This, again, confirms that $v$, from the perspective of the CS theory, is simply a free phenomenological parameter.
As a consequence, we can associate the parameters of the action $S_{2D}$ \eqref{3.18} to physical quantities. In particular, by matching through \eqref{3.20} the algebra \eqref{3.1}, written in terms of the boundary field $\Phi(X)$, with the algebra \eqref{3.21} written in terms of the density operator $n(t,\theta)$, we obtain the well known relation between the filling factor $\nu$ and the coupling constant of the CS theory $\kappa$ \cite{Blasi:2008gt,Blasi:2011pf}
\begin{equation}
\kappa=\frac{1}{2\pi\nu}\;.
\label{3.22}\end{equation}
Hence, from \eqref{3.16}, we can identify $v(t,\theta)$ as the spacetime-dependent velocity of the chiral boson $\Phi(X)$.
Notice that, because of \eqref{3.17}, the chiral boson turn left or right depending on the sign of $a_2$ in \eqref{3.18}, being $\tilde\kappa$ positive due to the positivity of the central charge. This is in agreement with \cite{Wen:1990se}, where the edge excitations were studied 
on a disc and on a cylinder, $i.e.$ on curved boundaries in flat spacetime, which belong to the cases studied in this paper. It is interesting to remark that we recover those results, in particular the algebra and the Luttinger theory, in a quite different way. The Luttinger liquids considered in \cite{Wen:1990se} on curved boundaries contain both right moving and left moving excitations, which corresponds to the possibility we mentioned of having a both left and right chiral boson.

\section{Summary of results and discussion}\label{sec4}

The aim of this paper is to understand if the geometry of the bulk spacetime affects in some way the boundary physics of CS theory, which, in the flat case, is known to reproduce on the boundary the theory of edge states of the Fractional Quantum Hall Effect. But if and how the details of the bulk metric have consequences on the physical observables has not been clarified yet. Naively, starting from a TFT one would expect a mild dependence on the bulk metric. Instead, our analysis shows that 
the velocity of the edge chiral boson indirectly depends on the bulk metric, through the determinant of the induced metric. Our work might therefore be the first step toward a theoretical framework for the recently observed accelerated chiral boson on the edge of particular Hall systems, which cannot be explained by the usual description in terms of flat CS theory with boundary. On the other hand, on the boundary of CS theory we find the usual KM algebraic structure of the flat background case, with the same central charge proportional to the inverse of the CS coupling. Therefore, the KM algebra appears to be insensitive to both the bulk metric and the type of boundary (at least, it is the same for planar and radial boundary). 
\\

A few remarks are in order.
\begin{itemize}
\item
The determinant of the metric depends on the coordinates that are chosen, and locally one might always find coordinates such that the determinant is a constant. Moreover, in 2+1 dimensions the boundary of CS theory is well known to be a CFT, hence depends only upon the conformal class of the boundary metric, and in every conformal class there is a metric of constant determinant. This would correspond to choosing a reference frame where the velocity of the chiral boson is normalized at a given constant value. This represents a good, physical reason for not choosing a metric with constant determinant. Starting with a not necessarily flat bulk Lorentzian metric, we recover on the boundary of CS theory two well known results: the KM algebra \eqref{2.43} and the Luttinger theory \eqref{3.18}. The boundary conformal structure is ensured by the existence of an algebraic structure of the KM type. Had we finished here, we would not have a good reason for not choosing, amongst the metrics in GNC, those with constant determinant. This would have been a without loss of generality choice, and we would have achieved our result as far as only the conformal structure described by the boundary algebra is concerned. But this is only half of our results: on the $planar$ boundary of CS theory with $flat$ bulk metric it is known to exist the Luttinger theory of the chiral boson. We derived the corresponding action 
following the procedure described in Section 3.2. The action \eqref{3.18} is not invariant under change of coordinates, which hence is not allowed at this stage, and it was not evident, at least to us, to recover the same action, with the same non-covariant metric as the one emerging from a flat bulk spacetime. Nonetheless, the Luttinger theory which we found on the boundary of the CS theory with non-flat metric is not 
independent from its bulk counterpart. We would not have found this result if, in the previous step, we had frozen the bulk metric by choosing coordinates with constant determinant (or a reference frame where the chiral velocity is constant). Instead, a  memory of the bulk metric survives, in the velocity of the chiral boson, which depends only on the determinant of the bulk metric. Not having made the choice of constant determinant leaves therefore the possibility of highlighting the residual dependence on the bulk metric, and also of selecting the bulk metric by means of a physical input, which is the velocity of the edge chiral bosons, which is a physical observable. This is not so different from what is usually done with the CS coupling constant, which is fixed by means of the physical requirements of 
the incompressible Hall fluid, by normalizing the external current coupled to the electromagnetic gauge field to the known relationship between CS coupling constant and filling factor $\kappa=\frac{1}{2\pi\nu}$. Yet, speaking of a coupling constant for the $abelian$ CS theory is rather inappropriate, since it is always possible to reabsorb the coupling constant by means of a redefinition of the gauge fields. Only in the non-abelian case a true coupling constant for CS theory exists. Nevertheless, normalizing to one the CS coupling constant would be equivalent to loosing the entire structure of Landau levels of (F)QHE, and for that reason we keep it alive. We see an analogy between the two situations: as we do not normalize to one the CS coupling constant in the abelian case in order to describe the filling factors of the FQHE, for the same reason we do not choose coordinates such that the determinant is constant, in order to keep the possibility of fixing the bulk metric by measuring the (possibly time dependent) velocity of the edge chiral boson.
\item
The literature on CS theory, and, more in general, on TFT with boundary, is huge, and a comprehensive list of references on the subject is out of reach. 
Nevertheless, we did not find any derivation of the measurable quantities characterizing the Hall systems, like in particular the edge chiral velocities, in non-flat backgrounds. 
One reason might be the lack of experimental motivations: in most cases, the edge chiral velocities are constant, and these are already well explained by flat CS theory with planar boundary. However, similar results to the ones achieved in this paper can be found in \cite{Geiller:2017xad}, where the boundary degrees of freedom of Abelian Chern-Simons theory are studied on a generic manifold, implementing, in a way similar to what we did, the boundary condition by means of a lower dimensional term in the action. The 2D Floreanini-Jackiw action is found, coinciding with \eqref{3.18}, which leads to the equation for a chiral boson, but with constant chiral velocity. An interesting relation to the abelian Wess-Zumino-Novikov-Witten theory for a scalar field coupled to the gauge field \cite{Carlip:1994gy} is suggested, which certainly deserves careful attention.
\item
One of the main results of this paper is that the measurable chiral velocity $v$ of the edge modes is related not only to the CS level $\kappa$, as it is known already for a flat background, but also to the determinant of the metric, through which $v$ can acquire a time dependence, which can be measured.
It is indeed possible to detect time dependent edge chiral velocity, as well as the investigation of non-standard geometries (for what concerns the QHE). In \cite{Bocquillon} the boundary of a Hall system with $\nu=2$ is considered. The two channels (left and right) interact, and the problem diagonalizes into one fast and one slow mode, and the velocity is not constant and is measured. For what concerns non conventional bulk geometries, recently, in \cite{kumar},
the FQHE has been observed in graphene. Here, the electrons live on a two-dimensional ``suspended'' membrane made by carbon atoms. There are some differences with respect to the standard case (Dirac-type linear dispersion relation rather than quadratic, presence of an additional degree of freedom like a pseudo-spin, for instance), which makes the phenomenology similar to that of FQHE, but with different values of the plateaux. The nice thing is that graphene technically allows to make samples with various shapes like Corbino discs,  and observations seem sensitive to the bulk geometry. Moreover, always using graphene, there are attempts to realize QH systems in curved space, like described in 
\cite{wagner} and in \cite{Can:2014ota}. Now, we do not know if the results presented in this paper can immediately be applied to these particular experimental sets, but the remarkable fact is that a recent experimental research activity exists, which concerns the main topic of our paper: the possibility of having boundary chiral modes whose time dependent velocities are sensitive to the bulk geometry. 
\item
It is useful to make a comparison with the approach usually adopted in the Literature to describe non-constant velocities of the edge excitations in the Hall systems. It is known that the fermions in a quantum Hall droplet are constrained by a confining potential whose gradient determines the velocity of the chiral edge modes \cite{Kane95}. For instance, the confinement may involve potentials that depend on space and/or time due to deformations or interactions in the sample \cite{Hashi18} and typically, in the phenomenological framework, one can invoke these considerations in order to take into account non-constant chiral velocities, which therefore, in the microscopic model, depend on the different possible potentials in the sample. 
In general, the paradigmatic way to describe the edge excitations has been illustrated by Wen \cite{Wen:1989mw,Wen:1990qp,Wen:1991ty}, and well summarized in \cite{Kane95}. One starts with a 3D Chern-Simons action in $flat$ spacetime, from which the bulk DOF are eliminated by integrating out the time-component of the gauge field $A_0(x)$. Then, the irrotational constraint on the spatial components of the gauge field is $imposed$: $\vec\nabla\times \vec A = 0$. The boundary DOF is then identified by the scalar field characterizing the solution $\vec A = \vec\nabla\Phi$. Finally, the ``appropriate effective action'' at the edge is written as the sum of two terms:  the Luttinger action, describing a chiral boson with $constant$ velocity, to which is possibly added an interaction term, which depends on the form of the edge confining potential and on the details of the electron-electron interactions. If this latter interaction is assumed to be local, this obviously implies non-constant velocities. Usually, to simplify the description, the interactions are taken as piecewise functions, in order to take into account, for instance, the screening effects which are present in the leads. More details can be found in \cite{safi} and in \cite{perfetto}. This latter paper is related to an interesting experiment \cite{naturenanotech}. Other experiments are described in \cite{brasseur,lin}.
To make it short, the total interaction is split into two parts: a constant velocity piece $v$ contained in the Luttinger action and an additional local velocity term in the interaction. This is the framework for the description of edge excitations with non-constant velocities. Alternatively, according to our approach, there is no need of $imposing$ the irrotational condition which gives rise to the scalar boundary degree of freedom. Rather, it emerges as the breaking term of the Ward identity \eqref{2.25}, evaluated on shell. Similarly, the 2D action \eqref{3.18} is not $chosen$ as an effective action, but is recovered by means of a general procedure which we described as an ``holografic contact'' between the boundary conditions on the bulk gauge field and the EOM of the 2D scalar field. Finally, we showed that a bulk CS model on curved spacetime yields a possible $non$-$constant$ chiral velocity, where the locality property is inherited solely from the induced metric determinant from a bulk-boundary correspondence.
An intriguing way to interpret this fact, in relation to the microscopic models, is to consider it as an equivalent description: a change of potential can be effectively encoded in the CS theory by a bulk metric. As the CS coupling constant $\kappa$ encodes all the possible values of the filling factor $\nu$, similarly the CS bulk metric encodes the spacetime properties of the (confining) potential, in terms of the induced metric determinant.
What is certainly true is that the CS bulk theory in flat spacetime, when considered with a boundary, generates local DOF on the 2D induced theory,  corresponding to the edge states of FQHE, with $constant$ chiral velocities. The possibility of local velocities is not captured by the flat approach alone. 
\end{itemize}

\section*{Acknowledgments}

We thank Nicola Pinamonti and Alberto Blasi for enlightening discussions. We are particularly grateful to Dario Ferraro and Maura Sassetti for informing us about the phenomenological and experimental scenario related to the content of this paper. This work has been partially supported by the INFN Scientific Initiative GSS: ``Gauge Theory, Strings and Supergravity''. E.B. is supported by MIUR grant ``Dipartimenti di Eccellenza” (100020-2018-SD-DIP-ECC\_001).

\appendix

\section{Derivative of the Heaviside step distribution} \label{appA}

The Heaviside theta distribution is used to describe a boundary. Its general form is $\theta(f(x))$, where $f(x)=f(t,r,\theta)=0$ describes the equation of the hypersurface $\partial M$ of the manifold $M$, $i.e.$
\begin{equation}\label{A.1}
\theta(f(x))= \begin{cases} 1\;, & \mbox{if } x\in M \\ 0\;, & \mbox{if } x\not\in M \end{cases}\ .
\end{equation}
On the other side, Stokes theorem states that if the $n$-dimensional manifold $M$ has a boundary, then
\begin{equation}\label{A.2}
\int_Md^nx\; \sqrt{|g|}\, \nabla_\mu V^\mu 
=
\int_{\partial M} d^{n-1}y\; \sqrt{|\gamma|}\; n_\mu V^\mu\ ,
\end{equation}
where $n^\mu$ is the vector normal to the boundary and $\gamma$ is the determinant of the induced metric. The presence of the boundary $f(x)=0$ in the l.h.s.  of \eqref{A.2} can be implemented through the introduction of the step distribution \eqref{A.1}~:
	\begin{empheq}{align}\label{A.3}
	\int_Md^nx\sqrt{|g|}\ \nabla_\mu V^\mu&=\int d^nx\sqrt{|g|}\ \theta(f(x))\nabla_\mu V^\mu\\
	&=-\int d^nx\sqrt{|g|}\ \nabla_\mu\theta(f(x)) V^\mu\ .\nonumber
	\end{empheq}
The integration on the r.h.s.  of \eqref{A.3} is performed over all spacetime ($i.e.$ boundary at infinity), then, when integrating by parts on the second line, the boundary term vanishes. We can also observe, from \eqref{A.3}, that the step function is a scalar quantity. Identifying the r.h.s.  of \eqref{A.2} and \eqref{A.3} we get 	\begin{equation}\label{A.4}
	-\int d^nx\sqrt{|g|}\ \nabla_\mu\theta(f(x)) V^\mu=\int_{\partial M}d^{n-1}x\sqrt{|\gamma|}\ n_\mu V^\mu\ ,
	\end{equation}
where $n_\mu\equiv\pm\frac{\partial_\mu f(x)}{\sqrt{g^{\mu\nu}}\partial_\mu f\partial_\nu f}$ and the sign depends on the orientation of the hypersurface. In \eqref{A.4} we can identify $\nabla_\mu\theta(f(x))$ as a scalar Dirac delta $\delta_{\partial M}$\footnote{We adopted the notation of \cite{Vassilevich:2004id}, p.13 Eq.(4.10). }
	\begin{equation}\label{A.5}
	-n_\mu\;\delta_{\partial M}\equiv\nabla_\mu\theta(f(x))\ .
	\end{equation}
To see it explicitly and deduce how this distribution acts, we insert \eqref{A.5} back  in \eqref{A.4}
	\begin{equation}\label{A.6}
	\int d^nx\sqrt{|g|}\;\delta_{\partial M}\;n_\mu V^\mu=\int_{\partial M}d^{n-1}x\sqrt{|\gamma|}\ n_\mu V^\mu\ ,
	\end{equation}
and since in the case we consider in this paper the boundary is at $f(r)=R-r=0$, and by virtue of the GNC \eqref{1.11}, we have
\begin{equation}\label{A.7}
\delta_{\partial M}=\frac{\sqrt{|\gamma|}}{\sqrt{|g|}}\delta(R-r)=\delta(R-r)\ .
\end{equation}
Finally from \eqref{A.5}, \eqref{A.7} and, for this boundary in GNC \eqref{1.11},
\begin{equation}\label{A.8}
n_\mu=\delta_\mu^r\ ,
\end{equation}
we deduce
\begin{equation}
\nabla_\mu\theta(R-r)=-\delta_\mu^r\delta(R-r)\ .
\label{A.9}\end{equation}



\medskip

\begin{thebibliography}{15}
\bibitem{Witten:1988ze}
E.~Witten,
Commun. Math. Phys. \textbf{117} (1988) 353, doi:10.1007/BF01223371.
\bibitem{Witten:1988hf}
E.~Witten,
Commun. Math. Phys. \textbf{121} (1989) 351-399,
doi:10.1007/BF01217730.
\bibitem{Moore:1989yh}
G.~W.~Moore and N.~Seiberg,
Phys.\ Lett.\ B \textbf{220} 422-430 (1989),
doi:10.1016/0370-2693(89)90897-6.
\bibitem{Wen:1992vi}
X.~G.~Wen,
Int. J. Mod. Phys. B \textbf{6} (1992) 1711-1762,
doi:10.1142/S0217979292000840.
\bibitem{Stone:1990iw}
M.~Stone,
Annals Phys. \textbf{207} (1991) 38-52,
doi:10.1016/0003-4916(91)90177-A.
\bibitem{Frohlich:1994vq}
J.~Frohlich, A.~H.~Chamseddine, F.~Gabbiani, T.~Kerler, C.~King, P.~A.~Marchetti, U.~M.~Studer and E.~Thiran,
``The Fractional quantum Hall effect, Chern-Simons theory, and integral lattices,''
In: Chatterji S.D. (eds) Proceedings of the International Congress of Mathematicians. Birkh\"auser, Basel. ETH-TH-94-18.
\bibitem{Bieri:2010za}
S.~Bieri and J.~Frohlich,
Comptes Rendus Physique \textbf{12} (2011), 332-346,
doi:10.1016/j.crhy.2011.02.001.
\bibitem{Froehlich:2018oce}
J.~Fr\"ohlich,
Rev. Math. Phys. \textbf{30} (2018) no.06, 1840007
doi:10.1142/9789813233867\_0013.
\bibitem{Hasan:2010xy}
M.~Z.~Hasan and C.~L.~Kane,
Rev. Mod. Phys. \textbf{82} (2010) 3045,
doi:10.1103/RevModPhys.82.3045.
\bibitem{Qi:2011zya}
X.~L.~Qi and S.~C.~Zhang,
``Topological insulators and superconductors,''
Rev. Mod. Phys. \textbf{83} (2011) no.4 1057-1110,
doi:10.1103/RevModPhys.83.1057.
\bibitem{Hasan:2010hm}
M.~Z.~Hasan and J.~E.~Moore,
Ann. Rev. Condensed Matter Phys. \textbf{2} (2011) 55-78,
doi:10.1146/annurev-conmatphys-062910-140432.
\bibitem{moorenature}
J.~E.~Moore, 
Nature vol. 464,7286 (2010): 194-8,
doi:10.1038/nature08916.
\bibitem{Aratyn:1984jz}
H.~Aratyn,
Phys. Rev. D \textbf{28} (1983) 2016-2018,
doi:10.1103/PhysRevD.28.2016.
\bibitem{Amoretti:2013xya}
A.~Amoretti, A.~Braggio, G.~Caruso, N.~Maggiore and N.~Magnoli,
Adv. High Energy Phys. \textbf{2014} (2014) 635286,
doi:10.1155/2014/635286.
\bibitem{Zhang:1992eu}
S.~C.~Zhang,
Int. J. Mod. Phys. B \textbf{6} (1992) 25-58,
doi:10.1142/S0217979292000037.
\bibitem{Kac:1967jr}
V.~Ka\c{c},   
Funct. Anal. Appl. \textbf{1} (1967), 328.
\bibitem{Moody:1966gf}
R.~Moody,
Bull. Am. Math. Soc. \textbf{73} (1967) 217-221,
doi:10.1090/S0002-9904-1967-11688-4.
\bibitem{Blasi:1990jq}
A.~Blasi and R.~Collina,
Phys. Lett. B \textbf{243} (1990) 99-104,
doi:10.1016/0370-2693(90)90963-7.
\bibitem{Emery:1991tf}
S.~Emery and O.~Piguet,
Helv. Phys. Acta \textbf{64} (1991), 1256-1270.
\bibitem{Balachandran:1991dw}
A.~P.~Balachandran, G.~Bimonte, K.~S.~Gupta and A.~Stern,
Int. J. Mod. Phys. A \textbf{7} (1992) 4655-4670,
doi:10.1142/S0217751X92002106.
\bibitem{Horowitz:1989ng}
G.~T.~Horowitz,
Commun.\ Math.\ Phys.\  \textbf{125} 417 (1989),
doi:10.1007/BF01218410.
\bibitem{Karlhede:1989hz}
A.~Karlhede and M.~Rocek,
Phys.\ Lett.\ B \textbf{224} 58-60 (1989),
doi:10.1016/0370-2693(89)91050-2.
\bibitem{Birmingham:1991ty}
D.~Birmingham, M.~Blau, M.~Rakowski and G.~Thompson,
Phys.\ Rept.\  \textbf{209} 129-340 (1991),
doi:10.1016/0370-1573(91)90117-5.
\bibitem{Blasi:2005vf}
A.~Blasi, N.~Maggiore and M.~Montobbio,
Nucl.\ Phys.\ B \textbf{740} 281-296 (2006),
doi:10.1016/j.nuclphysb.2006.01.028.
\bibitem{Blasi:2019wpq}
A.~Blasi and N.~Maggiore,
Symmetry \textbf{11} (2019) 921,
doi:10.3390/sym11070921.
\bibitem{Cho:2010rk}
G.~Y.~Cho and J.~E.~Moore,
Annals Phys. \textbf{326} (2011) 1515-1535,
doi:10.1016/j.aop.2010.12.011.
\bibitem{Balachandran:1992qg}
A.~P.~Balachandran and P.~Teotonio-Sobrinho,
Int. J. Mod. Phys. A \textbf{8} (1993) 723-752,
doi:10.1142/S0217751X9300028X.
\bibitem{Chen:2015gma}
X.~Chen, A.~Tiwari and S.~Ryu,
Phys. Rev. B \textbf{94} (2016) n.4 045113,
doi:10.1103/PhysRevB.94.045113.
\bibitem{Geiller:2019bti}
M.~Geiller and P.~Jai-akson,
JHEP \textbf{09} (2020) 134,
doi:10.1007/JHEP09(2020)134.
\bibitem{Tiwari:2017wqf}
A.~Tiwari, X.~Chen, K.~Shiozaki and S.~Ryu,
Phys. Rev. B \textbf{97} (2018) n.24 245133,
doi:10.1103/PhysRevB.97.245133.
\bibitem{Symanzik:1981wd}
K.~Symanzik,
``Schrodinger Representation and Casimir Effect in Renormalizable Quantum Field Theory,''
Nucl.\ Phys.\ B \textbf{190} 1-44 (1981),
doi:10.1016/0550-3213(81)90482-X.
\bibitem{Amoretti:2014iza}
A.~Amoretti, A.~Braggio, G.~Caruso, N.~Maggiore and N.~Magnoli,
Phys. Rev. D \textbf{90} (2014) n.12 125006,
doi:10.1103/PhysRevD.90.125006.
\bibitem{Amoretti:2013nv}
A.~Amoretti, A.~Blasi, G.~Caruso, N.~Maggiore and N.~Magnoli,
Eur. Phys. J. C \textbf{73} (2013) n.6 2461,
doi:10.1140/epjc/s10052-013-2461-3.
\bibitem{Maggiore:2018bxr}
N.~Maggiore,
Eur. Phys. J. Plus \textbf{133} (2018) n.7 281,
doi:10.1140/epjp/i2018-12130-y.
\bibitem{Maggiore:2019wie}
N.~Maggiore,
J. Phys. A \textbf{52} (2019) n.11 115401,
doi:10.1088/1751-8121/ab045a.
\bibitem{Bertolini:2020hgr}
E.~Bertolini and N.~Maggiore,
Symmetry \textbf{12} (2020) n.7 1134,
doi:10.3390/sym12071134.
\bibitem{Amoretti:2017xto}
A.~Amoretti, A.~Braggio, N.~Maggiore and N.~Magnoli,
Adv. Phys. X \textbf{2} (2017) n.2 409-427,
doi:10.1080/23746149.2017.1300509.
\bibitem{Amoretti:2014kba}
A.~Amoretti, A.~Braggio, G.~Caruso, N.~Maggiore and N.~Magnoli,
JHEP \textbf{04} (2014) 142,
doi:10.1007/JHEP04(2014)142.
\bibitem{Dunne:1998qy}
Dunne G.V. (1999) Aspects Of Chern-Simons Theory. In: Comtet A., Jolicoeur T., Ouvry S., David F. (eds) Aspects topologiques de la physique en basse dimension. Topological aspects of low dimensional systems. Les Houches - \'Ecole d'\'Et\'e de Physique Th\'eorique, vol 69. Springer, Berlin, Heidelberg. doi.org/10.1007/3-540-46637-1-3
\bibitem{Poisson:2011nh}
E.~Poisson, A.~Pound and I.~Vega,
Living Rev. Rel. \textbf{14} (2011) 7,
doi:10.12942/lrr-2011-7.
\bibitem{wald}
R.~M.~Wald, 
``General Relativity",
1984 University of Chicago Press, ISBN 9780226870335.
\bibitem{weinberg}
S.~Weinberg,
``Gravitation And Cosmology",
2008 Wiley, ISBN 9788126517558.
\bibitem{Carroll:2004st}
S.~Carroll,
``Spacetime and Geometry: An Introduction to General Relativity'', 2019,
Cambridge University Press,
ISBN 978-0-8053-8732-2, 978-1-108-48839-6, 978-1-108-77555-7.
\bibitem{d'inverno}
R.~d'Inverno, 
``Introducing Einstein's Relativity",
1990 Clarendon Press, ISBN: 9780198596868.
\bibitem{Nakanishi:1966zz}
N.~Nakanishi,
Prog.\ Theor.\ Phys.\  \textbf{35} 1111-1116 (1966),
doi:10.1143/PTP.35.1111
\bibitem{Lautrup:1967zz}
B.~Lautrup,
Kong.\ Dan.\ Vid.\ Sel.\ Mat.\ Fys.\ Med.\  \textbf{35} n.11 (1967),
NORDITA-214
\bibitem{Elitzur:1989nr}
S.~Elitzur, G.~W.~Moore, A.~Schwimmer and N.~Seiberg,
Nucl. Phys. B \textbf{326} (1989), 108-134,
doi:10.1016/0550-3213(89)90436-7.
\bibitem{Rubakov:2008nh}
V.~A.~Rubakov and P.~G.~Tinyakov,
Phys. Usp. \textbf{51} (2008), 759-792,
doi:10.1070/PU2008v051n08ABEH006600.
\bibitem{Blasi:2017pkk}
A.~Blasi and N.~Maggiore,
Eur. Phys. J. C \textbf{77} (2017) no.9, 614,
doi:10.1140/epjc/s10052-017-5205-y.
\bibitem{Blasi:2015lrg}
A.~Blasi and N.~Maggiore,
Class. Quant. Grav. \textbf{34} (2017) no.1, 015005
doi:10.1088/1361-6382/34/1/015005.
\bibitem{Geiller:2017xad}
M.~Geiller,
Nucl. Phys. B \textbf{924} (2017), 312-365,
doi:10.1016/j.nuclphysb.2017.09.010.
\bibitem{Maggiore:2017vjf}
N.~Maggiore,
Int. J. Mod. Phys. A \textbf{33} (2018) n.02 1850013,
doi:10.1142/S0217751X18500136.
\bibitem{Nash:1983cq}
C.~Nash and S.~Sen,
``Topology and geometry for physicists.''
Academic Press, 1988, ISBN 9780080570853.
\bibitem{Warner}
F.W.~Warner,
``Foundations of Differentiable Manifolds and Lie Groups."
Springer Verlag, 1983, ISBN 9780387908946.
\bibitem{mack}
G.~Mack, ``Introduction to conformal invariant quantum field theory in two and more dimensions'', Carg\`ese Lectures July 1987, ed. by G. 't Hooft et al., Nonperturbative Quantum Field Theory, Plenum Press, N Y 1988, ISBN 978146130729-7.
\bibitem{Becchi:1988nh}
C.~Becchi and O.~Piguet,
Nucl. Phys. B \textbf{315} 153-165 (1989),
doi:10.1016/0550-3213(89)90452-5.
\bibitem{Basler:1991st}
M.~Basler,
Fortsch. Phys. \textbf{41} (1993), 1-43.
\bibitem{Wen:1989mw}
X.~G.~Wen,
Phys. Rev. B \textbf{43} (1991), 11025-11036,
doi:10.1103/PhysRevB.43.11025.
\bibitem{Wen:1990qp}
X.~G.~Wen,
Phys. Rev. Lett. \textbf{64} (1990), 2206,
doi:10.1103/PhysRevLett.64.2206.
\bibitem{Frohlich:1990xz}
J.~Frohlich and T.~Kerler,
Nucl. Phys. B \textbf{354} (1991), 369-417,
doi:10.1016/0550-3213(91)90360-A.
\bibitem{Haldane:1981zza}
F.~D.~M.~Haldane,
J. Phys. C \textbf{14} (1981) 2585-2609,
doi:10.1088/0022-3719/14/19/010.
\bibitem{Blasi:2008gt}
A.~Blasi, D.~Ferraro, N.~Maggiore, N.~Magnoli and M.~Sassetti,
Annalen Phys. \textbf{17} (2008) 885-896,
doi:10.1002/andp.200810323.
\bibitem{Blasi:2011pf}
A.~Blasi, A.~Braggio, M.~Carrega, D.~Ferraro, N.~Maggiore and N.~Magnoli,
New J. Phys. \textbf{14} (2012), 013060,
doi:10.1088/1367-2630/14/1/013060.
\bibitem{Wen:1990se}
X.~G.~Wen,
Phys. Rev. B \textbf{41} (1990), 12838-12844,
doi:10.1103/PhysRevB.41.12838.
\bibitem{Carlip:1994gy}
S.~Carlip,
Phys. Rev. D \textbf{51} (1995), 632-637
doi:10.1103/PhysRevD.51.632.
\bibitem{Bocquillon}
E.~Bocquillon,  V.~Freulon, J.M.~Berroir, P.~Degiovanni, B.~Pla\c{c}ais, A.~Cavanna, Y.~Jin and G.~F\`eve,
Nat Commun 4, 1839 (2013),
doi.org/10.1038/ncomms2788.
\bibitem{kumar}
M.~Kumar, A.~Laitinen and P.~ Hakonen, 
Nat Commun 9, 2776 (2018),
doi.org/10.1038/s41467-018-05094-8.
\bibitem{wagner}
G.~Wagner, F.~de Juan and D.~X.~Nguyen,
[arXiv:1911.02028 [cond-mat.str-el]].
\bibitem{Can:2014ota}
T.~Can, M.~Laskin and P.~Wiegmann,
Phys. Rev. Lett. \textbf{113} (2014), 046803,
doi:10.1103/PhysRevLett.113.046803.
\bibitem{Kane95}
C.~L.~Kane and M.~P.~Fisher,
Physical Rev. B, \textbf{51} (1995), 13449-13466,
doi: 10.1103/physrevb.51.13449.

\bibitem{Hashi18}
M.~Hashisaka and T.~Fujisawa,
Reviews in Physics, \textbf{3} (2018), 32-43,
doi: 10.1016/j.revip.2018.07.001.

\bibitem{Wen:1991ty}
X.~G.~Wen,
Phys. Rev. B \textbf{44} (1991) no.11, 5708,
doi:10.1103/PhysRevB.44.5708.

\bibitem{safi}
I.~Safi, 
Ann. Phys. Fr.  22 (5), 463-679,
doi.org/10.1051/anphys:199705001.

\bibitem{perfetto}
E.~Perfetto, G.~Stefanucci, H.~Kamata and T.~Fujisawa,
Phys. Rev. B 89, 201413(R),
doi.org/10.1103/PhysRevB.89.201413.

\bibitem{naturenanotech}
H.~Kamata, N.~Kumada, M.~Hashisaka, K.~Muraki and T.~Fujisawa,
Nature Nanotech 9, 177-181 (2014),
doi.org/10.1038/nnano.2013.312.

\bibitem{brasseur}
P.~Brasseur, N. H.~Tu, Y.~Sekine, K.~Muraki, M.~Hashisaka, T.~Fujisawa, and N.~Kumada,
Phys. Rev. B 96, 081101(R),
doi.org/10.1103/PhysRevB.96.081101.

\bibitem{lin}
C.~Lin, M.~Hashisaka, T.~Akiho, K.~Muraki, and T.~Fujisawa,
Phys. Rev. B 104, 125304,
doi.org/10.1103/PhysRevB.104.125304.

\bibitem{Vassilevich:2004id}
D.~V.~Vassilevich,
Contemp. Math. \textbf{366} (2005) 3-22.
\end{thebibliography}
\end{document}